# Spectrum Identification using a Dynamic Bayesian Network Model of Tandem Mass Spectra

**Ajit P. Singh**[†]    **John Halloran**[†]    **Jeff A. Bilmes**[†]    **Katrin Kirchoff**[†]    **William S. Noble**[‡]

[†]Department of Electrical Engineering    [‡]Department of Genome Sciences
University of Washington                University of Washington
Seattle, WA 98195                    Seattle, WA 98195

## Abstract

Shotgun proteomics is a high-throughput technology used to identify unknown proteins in a complex mixture. At the heart of this process is a prediction task, the *spectrum identification problem*, in which each fragmentation spectrum produced by a shotgun proteomics experiment must be mapped to the peptide (protein subsequence) which generated the spectrum. We propose a new algorithm for spectrum identification, based on dynamic Bayesian networks, which significantly outperforms the de-facto standard tools for this task: SEQUEST and Mascot.

## 1    Introduction

Shotgun proteomics is the dominant technology used to identify which proteins are present in a cell or tissue sample, and is widely used in the biological sciences (Marcotte, 2007; Steen and Mann, 2004). At the heart of shotgun proteomics is a machine learning problem. Proteins are broken down into small fragments, called peptides. Although shotgun proteomics cannot directly determine the sequence of these peptides, the technology can rapidly generate indirect information about peptide sequences, called fragmentation spectra. The task of identifying the peptide string responsible for generating an observed fragmentation spectrum is known as the *spectrum identification problem*. This problem is similar in form to speech recognition, in which a spoken utterance (fragmentation spectrum) is mapped to its corresponding natural language string (peptide sequence).

This paper applies dynamic Bayesian networks (DBNs) to the spectrum identification problem. We draw on ideas from the graphical models community to build an algorithm for spectrum identification which is significantly more accurate than the most popular tools in wet-lab use, SEQUEST (Eng et al., 1994) and Mascot (Perkins et al., 1999), as well as other representative tools which have been proposed for spectrum identification. We call our algorithm Didea[1].

## 2    Background

Spectrum identification is a machine learning problem, but like many problems in computational biology, one which has a high barrier to entry for computer scientists. Therefore, we present a self-contained introduction to shotgun proteomics, explaining how peptides generate fragmentation spectra (Section 2.1). One of the reasons for using a dynamic Bayesian network is that it allows us to use qualitative knowledge about the physics of peptide fragmentation (in the structure of the DBN) and a small number of trainable parameters to improve predictive power.

We review the literature on spectrum identification, which falls into two broad categories: database search (Section 2.2) and *de novo* identification (Section 2.3). Database search uses additional biological information about the sample being analyzed to constrain the statistical complexity of spectrum identification. Database search is more accurate than *de novo*, and is the preferred technique in practice (Kim et al., 2010). The two most popular tools in wet-lab practice, SEQUEST and Mascot, are both database search tools, as is Didea.

### 2.1    Shotgun Proteomics

A typical shotgun proteomics experiment proceeds in three steps, as illustrated in Figure 1(a). The input to the experiment is a collection of proteins, which have been isolated from a complex mixture. Each protein can be represented as a string of amino acids, where the alphabet is size 20 and the proteins range in length from 50–1500 amino acids. A typical complex mixture may

---

[1]A portmanteau of the words 'Dynamic', 'Peptide', and 'Algorithm'.

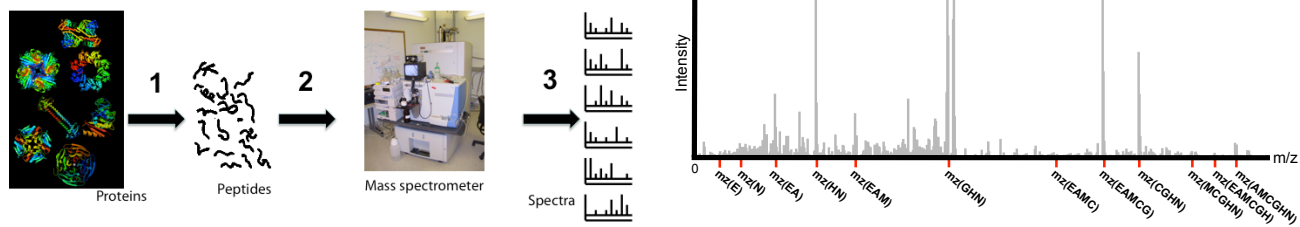

(a) The three steps—(1) cleaving proteins into peptides, (2) separation of the peptides using liquid chromatography, and (3) tandem mass spectrometry analysis.

(b) Red ticks indicate where we expect to detect ions from the fragmentation of peptide EAMCGHN.

Figure 1: The schematic for a typical shotgun proteomics experiment (a), with an example of a real fragmentation spectrum annotated by the theoretical spectrum, in red, of a candidate peptide EAMCGHN (b).

contain a few thousand proteins, ranging in abundance from tens to hundreds of thousands of copies.

In the first experimental step, the proteins are digested into peptides, or protein subsequences, using a molecular agent called trypsin. On average, the length of the peptides for the model organisms we consider is 14-15, with no peptides being longer than 50 amino acids. Digestion is necessary because whole proteins are too massive to be subject to direct mass spectrometry analysis without using very expensive equipment. In the second experimental step, peptides are subjected to a process called *liquid chromatography*, in which the peptides pass through a thin glass column that separates the mixture of peptides based on a particular chemical property (e.g., hydrophobicity). This separation step reduces the complexity of the mixtures of peptides going into a mass spectrometer. The third experimental step, which occurs inside the mass spectrometer, involves two rounds of mass spectrometry. Approximately every second, the device analyzes the population of approximately 20,000 intact peptides that most recently exited through the liquid chromatography column. Then, based on this initial analysis, the machine selects five distinct peptide species for fragmentation. Each of these fragmented species is subjected to a second round of mass spectrometry analysis. The resulting "fragmentation spectra" are the primary output of a shotgun proteomics experiment.

A fragmentation spectrum is shown in Figure 1(b). We explain how such a spectrum is generated using a concrete example. Assume that, from the population of 20,000 intact peptides, we isolate a distinct peptide species: a collection of peptide molecules, each with sequence EAMCGHN. Each peptide molecule has extra protons, which allows it to be isolated and accelerated using magnetic fields. Molecules with extra protons are called ions; the number of extra protons, its charge. These isolated peptide molecules are collided into a neutral gas, which causes each molecule to break, typically along one of the amino acid bonds at random. For a peptide of length $n$, there are $n-1$ bonds which can be broken, each yielding a (prefix, suffix) pair: e.g., (E, AMCGHN), (EA, MCGHN), ..., (EAMCGH, N). When an amino acid bond is broken, the extra protons migrate to either the prefix or suffix, at random. The charged prefix is called a *b*-ion; the charged suffix is called a *y*-ion. Collectively, the ionized products of fragmentation are called product ions. We assume all product ions are either *b*- or *y*-ions. The sequence of peptides or product ions cannot be directly measured, but we can measure how often product ions with a particular mass-to-charge (m/z) ratio are detected. These measurements are represented in a plot with m/z (measured in Daltons) on the horizontal axis and intensity (unitless, but roughly proportional to ion abundance) on the vertical axis.

Real fragmentation spectra are noisy. The isolation of peptide species is imperfect, so fragmentation spectra can contain product ions from peptides with different sequences (chimeric spectra). Mass analyzers can measure subatomic mass differences, so even the smallest unfiltered contaminant adds noise to the fragmentation spectrum. Product ions without charge cannot be detected, and we do not always know how many protons are adopted by a product ion—i.e., variable charge. There are a host of secondary fragmentations and degradations of product ions, which are measured. Finally, intensity measurements are quite noisy.

The input to the spectrum identification problem is a fragmentation spectrum, along with the observed (approximate) mass of the intact peptide whose fragmentation produced the spectrum. The output is the sequence of the unknown peptide, the fragmentation of which generated the spectrum. A peptide paired with a spectrum is called a peptide-spectrum match (PSM).

## 2.2 Database Search

A tandem mass spectrometry experiment produces a set of fragmentation spectra $S = \{s_1, \ldots, s_r\}$. Each spectrum is also associated with a measurement of the mass of the unknown peptide which generated the spectrum, $m(s_i)$. Let $U$ be the universe of all peptides. We assume as input a database of possible peptides, $P \subset U$. The key assumptions behind database search are that (i) we know the organisms from which the proteins came from, and (ii) we have a set of all known proteins for the organism, which can be computed given the organism's genome. We need only search over peptides in $P$, not all peptides in $U$. Formally, spectrum identification is the task of assigning a peptide $p_i \in P$ to each spectrum $s_i$. Let $\Psi : S \times P \to \mathbb{R}$ denote a scoring function, where a higher score corresponds to a higher confidence that a peptide-spectrum match is correct (i.e., the peptide generated the spectrum).

Since we have measured $m(s_i)$, we can further constrain the search space over peptides to those whose mass is close to $m(s_i)$, a set of candidate peptides,

$$C(s_i, P, \delta) = \{p : p \in P, |m(p) - m(s_i)| < \delta\}, \quad (1)$$

where $m(p)$ is the mass of the peptide $p$. In our experiments, we use $\delta = 3.0$ Daltons. The database search itself involves scoring all candidate peptides, returning the highest scoring one for each spectrum. For $i = 1 \ldots r$,

$$p_i = \underset{p \in C(s_i, P, \delta)}{\operatorname{argmax}} \ \Psi(s_i, p). \quad (2)$$

The first computer program to use a database search procedure to identify fragmentation spectra was SE-QUEST (Eng et al., 1994), whose scoring function is defined in terms of inner products between the quantized observed spectrum and a theoretical spectrum $\phi(p)$ generated from a simple model of peptide fragmentation. Observed and theoretical spectra are vectors of real numbers, so a peptide can be compared to a spectrum using inner products,

$$\text{XCorr}(s, \phi(p)) = \alpha_x - \beta_x, \quad \text{where}$$

$$\alpha_x = \langle s, \phi(p) \rangle, \ \ \beta_x = \frac{1}{150} \sum_{\tau = -75}^{75} \langle s, \phi_\tau(p) \rangle, \quad (3)$$

and where $\phi_\tau(p)$ is just $\phi(p)$ shifted by $|\tau|$ units to the right ($\tau > 0$) or left ($\tau < 0$). Intuitively, a peptide is a good match if the observed spectrum $s$ is highly correlated to the theoretical spectrum ($\alpha_x$ high), but not to shifted versions of the theoretical spectrum ($\beta_x$ low). Didea's scoring function has an analogous $\alpha - \beta$ representation, where the analogues of correlation ($\alpha$) and cross-correlation ($\beta$) are calculated using graphical model inference (Section 3.2).

There are a number of different tools for database search: e.g., SEQUEST (Eng et al., 1994), Mascot (Perkins et al., 1999), OMMSA (Geer et al., 2004), X!Tandem (Craig and Beavis, 2004), PepHMM (Wan et al., 2006), Riptide (Klammer et al., 2008), Andromeda (Cox et al., 2011), InsPecT (Tanner et al., 2005) and MS-GFDB (Kim et al., 2010). Separate from these approaches are post-processors, such as Peptide-Prophet (Keller et al., 2002) and Percolator (Käll et al., 2007), which refine the scores of a set of prescored peptide-spectrum matches, instead of scoring spectra in a streaming fashion. We note that statistical tools also exist for identifying proteins from peptide identifications, which can also refine peptide scores like a post-processor (e.g., (Li et al., 2010)). Some of the tools listed do not have implementations available for benchmarking (e.g., Riptide). Indeed, each of the above papers supports its predictive power claims by benchmarking against a representative subset of prior work. We have opted for a large set of competitors, on a diverse panel of data sets: SEQUEST, Mascot, MS-GFDB, OMMSA. SEQUEST and Mascot are the two most popular tools in wet-lab practice: e.g., each has been cited over 3000 times in Google Scholar, and both are commercial products.[2] MS-GFDB is a new scoring tool; the authors have demonstrated superior performance on the same kinds of fragmentation spectra that we study (collision-induced dissociation of tryptic peptides) against many of the tools listed above. OMMSA is open-source, and readily available. We do not compare against post-processors, as they use information not available to the other tools.

## 2.3 De Novo vs. Database Search

Early work on spectrum identification framed the problem as an exhaustive search over all possible peptides (Sakurai et al., 1984), an approach which came to be known as *de novo* spectrum identification, a.k.a. de novo peptide sequencing (Bafna and Edwards, 2003; Bandeira et al., 2008; Bartels, 1990; Bern and Goldberg, 2005; Bhatia et al., 2011; Dancik et al., 1999a,b; Datta and Bern, 2008; Fischer et al., 2004; Frank and Pevzner, 2005; Frank et al., 2005; Jeong et al., 2010). De novo tools select all candidate peptides that are within a mass tolerance of $m(s_i)$, not just those peptides that occur in a per-organism peptide database $P$.

De novo and database search methods cannot be equitably compared. Database search uses knowledge about the organism from which the protein sample is drawn, as well as knowledge of the proteome of that organism. De novo does not assume that such information is available. De novo searches over pep-

---

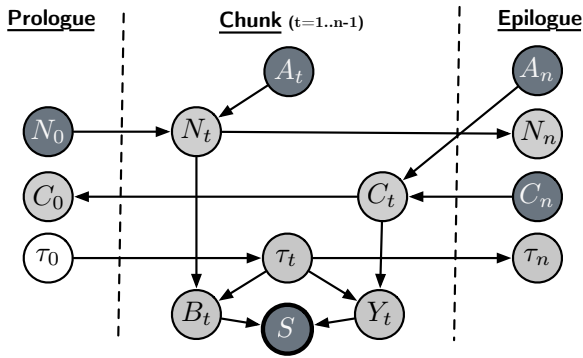

**Prologue**  **Chunk** (t=1..n-1)  **Epilogue**

Figure 2: Basic Didea model. Dark gray nodes are observed random variables, light gray nodes are variables whose value is a deterministic function of its parents, and unshaded nodes are hidden random variables. Bold circles indicate vectors of random variables.

tides in $C(s_i, U, \delta)$; database search searches over peptides in $C(s_i, P, \delta)$, where for any reasonable value of $\delta$, $|C(s_i, P, \delta)| \ll |C(s_i, U, \delta)|$.

## 3 Generative Model of Peptide Fragmentation

Didea encodes much of the physical process of peptide fragmentation by collision-induced dissociation within a dynamic Bayesian network. To maximize predictive performance, we take advantage of the idea of spectrum shifting (Equation 3) within our model, to derive a scoring function for peptide-spectrum matches based on inference within the DBN. The resulting inference-based scoring function can be viewed as an $\alpha - \beta$ score, where instead of dot products, we use the DBN to induce analogues of correlation and cross-correlation.

Figure 2 shows the basic Didea model; Sections 3.1–3.3 describe all the conditional probability distributions. Many of the conditional probability distributions are deterministic, which allows the scoring function to be computed in polynomial time.

An advantage of Didea is that adding more knowledge to the model (e.g., Sections 3.4-3.5) yields a new scoring function without any changes to the code. We just compute the same posterior on a new model.

### 3.1 Mapping the peptide to theoretical product ions

A DBN is a structured probability distribution over variable-length sequences of random variables, and is a strict generalization of the more familiar hidden Markov model (HMM). In Didea, a peptide containing $n$ amino acids is represented by a sequence of random variables

($A_t : t = 1 \dots n$). Each variable $A_t$ takes on the value of one of the 20 standard amino acids. In database search the peptide is given, so the amino acid variables are observed: $\mathbf{p} = (A_t = a_t : t = 1 \dots n)$. To simplify the main text, we equivalently refer to a peptide as a sequence of amino acids $p = a_1 a_2 \dots a_n$.

In collision-induced dissociation, most of the product ions are either $b$- or $y$-ions. If the peptide $p$ is cleaved between $a_t$ and $a_{t+1}$, then the resulting (respectively $b$- and $y$-) product ions are referred to as $b_t = a_1 \dots a_t$ and $y_t = a_{t+1} \dots a_n$. We use the function $c(\cdot)$ to denote the charge of a peptide, $c(p)$; the charge of the product ions, $c(b_t)$ and $c(y_t)$; and the observed charge(s) of the fragmentation spectrum, $c(s) \subseteq \{+1, +2, +3\}$.[3] For now, we assume that $c(p) = +2$ (two adopted protons) and that the product ions each adopt one proton from the peptide during fragmentation, i.e., $\forall t, c(b_t) = c(y_t) = +1$. These charge assumptions simplify exposition, and are relaxed in Section 3.5.

The structure of the Didea model is shown in Figure 2. Random variables are grouped into *frames*, indexed by $t = 0 \dots n$. The first frame (frame 0) is referred to as the prologue; the last frame (frame $n$), the epilogue. The variables in the prologue and epilogue do not themselves correspond to $b$- or $y$-ions, but do contain variables needed to define recursions used to compute the mass of product ions. For a peptide of length $n$, the chunk frame is repeated once for $a_1, \dots, a_{n-1}$, in the same way that variables are repeated in an HMM to accommodate variable length sequences.

A product ion (a.k.a. fragmentation) spectrum $s$ is a collection of peaks $s = \{(x_j, h_j)\}_j$, where $x_j$ is a point on the mass-to-charge, or m/z, axis (x-axis), and $h_j$ is the corresponding intensity. To check whether the $b_t$ or $y_t$ ion appears in the observed spectrum, we need the mass of each product ion. The neutral masses of the $b_t$ and $y_t$ ions are represented by random variables $N_t \equiv m(A_1 \dots A_t)$ and $C_t \equiv m(A_{t+1} \dots A_n)$, where $m(\cdot)$ returns the neutral masses of a sequence of amino acids. In Didea, the mass of each amino acid is rounded to the nearest whole Dalton, so that $N_t$ and $C_t$ are multinomial random variables over the neutral mass of the $b_t$ and $y_t$ ions—$N_t, C_t \in \{0, \dots, D\} \subset \mathbb{Z}_{++}$. By definition, $N_0 = 0$ and $C_n = 0$, so that the masses of the product ions can be computed recursively: $N_t = N_{t-1} + m(A_t)$, $C_t = C_{t+1} + m(A_t)$. Note that $C_0 = N_n = m(A_1 \dots A_n)$. Unlike HMMs, DBNs can have both left-to-right and right-to-left arcs (as long as there are no directed cycles), which allows us to implement the two recursions in the same model. Since $\{A_t\}_t$ are observed, both $N_t$ and $C_t$ become deterministic

---

[3]In our experiments, the mass spectrometer operates in positive-ion mode, and the reported charge of each spectrum is either $c(s) = +1$, $c(s) = +2$, or $c(s) \in \{+2, +3\}$.

random variables: $p(N_t = m(a_1 \ldots a_t)) = 1$ and $p(C_t = m(a_{t+1} \ldots a_n)) = 1$.

## 3.2 Mapping theoretical product ions to the fragmentation spectrum

We now have the mass of each product ion, and we have assumed that the charge of each product ion is +1. To finish the basic Didea model, we need to *(i)* encode the spectrum as observed random variables, *(ii)* introduce random variables which map from mass and charge to a specific m/z region in the spectrum, and *(iii)* define a conditional probability distribution which measures the value of finding a peak of intensity $h$ in the m/z region where a product ion is expected to be.

From the settings used to collect our spectra, we know that $x_j \in [0, 2000]$ m/z units. For simplicity, we quantize the $m/z$ scanning range into bins that are 1Da wide. In Figure 2, the bins correspond to a vector of random variables $\mathbf{S} = (S_i : i = 1 \ldots B = 2000)$. A spectrum is an instantiation of the random variables $\mathbf{s} = (S_i = s_i : i = 1 \ldots B)$, where each $s_i \geq 0$.

Spectra may differ by orders of magnitude in both total intensity $(\sum_j h_j)$ and maximum intensity $(\max\{h_j\})$. To control for intensity variation, we rank-normalize each spectrum: peaks are sorted in order of increasing intensity, and the $i^{\text{th}}$ peak is assigned intensity $i/|s|$, so $\max\{h_j\} = 1.0$. If bin $i$ contains no peak, then $s_i = 0$. Otherwise, $s_i$ is the highest rank-normalized peak, so $s_i \in [0, 1]$. We represent the vector of random variables $\mathbf{S}$ using a single, bold-edged node in Figure 2.

Each repetition of the chunk frame represents one pair of $b$- and $y$-ions, denoted $(b_t, y_t)$. The random variables $B_t, Y_t \in \{1, \ldots B\}$ represent, respectively, the bin where the $b_t$ and $y_t$ ions are to be expected.[4] We want to be able to shift the theoretical spectrum, in much the same fashion as Equation 3. We do so using the discrete random variables $\tau_t \in [-M \ldots + M]$, for some $M \in \{1 \ldots B\}$ (we use $M = 37$). We expect to see a peak corresponding to the $b_t$ ion in $S_{B_t}$, and a peak corresponding to the $y_t$ ion in $S_{Y_t}$. The mass-to-$m/z$ mappings are $B_t = \text{round}(N_t + 1)$ and $Y_t = \text{round}(C_t + 19)$, where 1 is the mass of a proton, and 19 the mass of a proton plus a water molecule. $\tau_t$ shifts all the theoretical product ion peaks. To ensure that $B_t, Y_y \in [0, \ldots B]$, values less than 1 are changed to 1; values greater than $B$ are changed to $B$. To ensure that all the product ion peaks are shifted by the same amount, we copy the value of $\tau_0$ from the prologue frame into each subsequent one: $\forall t = 1 \ldots n$, $\tau_t = \tau_{t-1}$. As will be seen in Equation 6, to mimic the arithmetic average over shifts in the background term in XCorr, we assume that $\tau_0$ is uniformly distributed over all shifts, $p(\tau_0) = (2M + 1)^{-1}$.

Most of the conditional probability distributions in Figure 2 are deterministic, which leads to a concise form for the joint distribution. For now, we refer to the contribution of a spectrum peak to the likelihood as non-negative function $g$:

$$p(\tau_0, \mathbf{s}, \mathbf{p}) = p(\tau_0) \prod_{t=1}^{n-1} \prod_{i=1}^{B} g(S, i, b_t, y_t, \tau_t), \qquad (4)$$

$$g(\cdots) \triangleq P(S_i \,|\, b_t, y_t, \tau_t)^{\mathbf{1}(i = b_t + \tau_t \,\vee\, i = y_t + \tau_t)}. \qquad (5)$$

Equation 4 is the likelihood of the model in Figure 2.

The values of $B_t$ and $Y_t$, for each $t$, are deterministically determined from the peptide sequence. Assignment $\{B_t = b_t\}_t$ forces the model to assign some score, based on the intensity of the peaks found in $\{S_{b_t}\}_t$ (likewise with $\{Y_t = y_t\}_t$). The choice of $P(S_i \,|\, b_t, y_t, \tau_t)$ determines how the spectrum influences the score; details are deferred to Section 3.3.

One way to use the proposed model to score a PSM would be to set $\tau_0 = 0$ and use Equation 4, which would be analogous to dropping the $\beta_x$ term in XCorr. Instead, to achieve an analogue to XCorr, we propose a score that can be interpreted as the difference between the Didea analogues of spectrum correlation $(\alpha_d)$ and cross-correlation $(\beta_d)$:

$$\begin{aligned} \theta(\mathbf{s}, \mathbf{p}) &\triangleq \log p(\tau_0 = 0 \,|\, \mathbf{p}, \mathbf{s}) = \alpha_d - \beta_d, \\ \alpha_d &= \log p(\tau_0 = 0, \mathbf{p}, \mathbf{s}), \\ \beta_d &= \log p(\tau_0) \sum_{\tau_0} p(\mathbf{p}, \mathbf{s} \,|\, \tau_0). \end{aligned} \qquad (6)$$

Both $\alpha_x$ and $\alpha_d$ are biased, assigning higher scores to spectra that tend to be close matches to many peptides (e.g., spectra with more non-zero peaks will tend to have higher $\alpha$, regardless of which peptide is evaluated against it). The $\beta_d$ term, like the $\beta_x$ term, penalizes spectra which tend to match many peptides, which is why $\alpha - \beta$ scores tend to be more discriminative than $\alpha$. Equation 6 is a (posterior) probability; Equation 3 is not probabilistic.

## 3.3 Modeling spectrum peak intensity

Here, we define Equation 5, which determines the influence of the spectrum on the likelihood, and thus the PSM score $\theta(\mathbf{s}, \mathbf{p})$. The cost of inference is proportional to the number of times $P(S_i \,|\, b_t, y_t, \tau_t)$ is evaluated. Since the spectrum is fixed across all $t = 1 \ldots n - 1$, the values of $P(S_i \,|\, b_t, y_t, \tau_t)$, for all $(S_i = s_i, b_t, y_t, \tau_t)$, can be precomputed and stored in a lookup table:

---

[4]Since there is no ambiguity, we also use $b_t$ and $y_t$ to represent assignments to random variables, $B_t = b_t$ and $Y_t = y_t$.

$w : \{1 \dots B\} \to \mathbb{R}_+$, a transformation of the spectrum into non-negative weights. When a predicted product ion matches peak $s_i$ in the spectrum, $w(i)$ is the contribution of that peak match to the PSM score.

In a graphical model, any conditional probability distribution on discrete variables can be reparameterized into a lookup table using virtual evidence (Pearl, 1998), allowing us to re-express Equation 4 as

$$p(\tau_0, \mathbf{p}, \mathbf{s}) \propto p(\tau_0) \prod_{t=1}^{n-1} w(B_t + \tau_t) w(Y_t + \tau_t), \quad (7)$$

where $w(i) = f(s_i)$, for user-chosen transformation $f > 0$. Logically, we prefer to match theoretical product ions to higher intensity peaks in the observed spectrum. Our experiments with $f$ focused on scaled and shifted exponential distributions, restricted to $[0,1]$, since $s_i \in [0,1]$. Based on these experiments, a class of $f$ which performs well on a wide variety of data sets is

$$f_\lambda(S) = 1 - \lambda e^{-\lambda} + \lambda e^{-\lambda(1-S)}. \quad (8)$$

The parameter, $\lambda > 0$, dictates the value placed on matching higher intensity peaks in the scoring function. The larger $\lambda$ is, the higher the preference for matching $b$- and $y$-ions to high intensity peaks. There is no simple threshold for determining peak relevance based only on intensity; a PSM can have a high score even if most of the peaks identified are of low intensity. The unusual form of $f_\lambda$ came out of an attempt to use exponential distributions to model peak intensity. Since $f_\lambda(S)$ is nearly log-linear on $[0,1]$, one can instead use $f'_{\lambda_2}(S) = \exp(\lambda_2 S)$, where $\lambda_2 > 0$ is chosen to make $f'_{\lambda_2} \approx f_\lambda$. We get slightly better results using $f_\lambda$.

To optimize $\lambda$ with respect to $\theta(\mathbf{s}, \mathbf{p})$, we use a grid search on a development set of 1000 yeast spectra; $\lambda = 0.5$ yielded the best results. $\lambda$ acts as a tuning parameter for soft peak filtering. Both SEQUEST and MS-GFDB have to decide whether a peak is relevant before scoring, which can affect which PSM is selected.

### 3.4 Modeling fragmentation of charge +3 peptide ions

Thus far, we have assumed that the precursor ion has two protons, which are divided equally between product ions. However, electrospray ionization frequently produces peptide precursor ions with more than two protons (Trauger et al., 2002). That is, the charge of the spectrum is higher than +2. In this section, we extend Didea to model the fragmentation of charge +3 peptides.[5]

Here, we have that $c(p) = +3$, with $c(b_t)$ and $c(y_t)$ denoting the charges of the product ions. Since charge is conserved in fragmentation we have $\forall t$, $c(p) = c(b_t) + c(y_t)$. The charge of $b_t$ is modeled as a multinomial $\xi_t \in \{0, +1, +2, +3\}$. It is unusual for a product to consume all the protons, and uncharged products are not detectable. In the absence of prior knowledge, we assume that the remaining choices are equally probable, $p(\xi_t = +1) = p(\xi_t = +2) = 0.5$. Figure 3(a) is the model for charge +3 peptide ions.

The definition of the scoring function, Equation 6, remains unchanged. The inference still uses $\log p(\tau_n = 0 \mid \mathbf{p}, \mathbf{s})$ as the score of a PSM. However, the values of $\alpha_d$ and $\beta_d$ change, because the joint distribution over PSMs has changed:

$$p_{\mathrm{ch3}}(\tau_0, \mathbf{p}, \mathbf{s}) \propto p(\tau_0) \prod_{t=1}^{n-1} \sum_{\xi_t} p(\xi_t) \Omega(t), \quad (9)$$

$$\Omega(t) = \left[ w(B_t(N_t, \xi_t) + \tau_t) w(Y_t(C_t, \xi_t) + \tau_t) \right],$$

$$B_t(N_t, \xi_t) = \mathrm{round}\left( (N_t + \xi_t)/\xi_t \right), \quad (10)$$

$$Y_t(C_t, \xi_t) = \mathrm{round}\left( (C_t + 18 + \xi_t)/\xi_t \right). \quad (11)$$

Equations 10–11 are the mass to quantized m/z mappings for $b$- and $y$-ions. For each pair of ions $(b_t, y_t)$, inference integrates over all divisions of the charge.

SEQUEST has a very different treatment of charge-variants of product ions. The theoretical spectrum in SEQUEST consists of the union of peaks where $c(b_t) = +1$ and $c(b_t) = +2$. As a thought experiment, consider what peaks need to be in the observed spectrum to maximize the SEQUEST score. The $\alpha_x$ term is maximized when all the theoretical peaks appear in the spectrum; the $\beta_x$ term is minimized when no shifted version of the spectrum matches the theoretical peaks. To maximize the SEQUEST score, the observed spectrum would have to contain all peaks corresponding to all charge variants of the product ions, except those that would contribute significantly to the $\beta_x$ term. There can be an exponentially large gap between the number of peaks in the SEQUEST theoretical spectrum, and the observed spectrum. In contrast, our approach averages the score over all $2^{n-1}$ assignments to $\xi = (\xi_1 \dots \xi_{n-1})$.[6] Each assignment to $\xi$ corresponds to a particular combination of product ion peaks. Maximizing the score with respect to a peptide does not make the assumption that all charge-variants of the

---

[5] It is possible for electrospray ionization of tryptic peptides to produce peptide ions with a higher charge than +3,

---

though doing so depends both on the mass spectrometer, and properties of the peptide (Kinter and Sherman, 2000, p. 69). Our data does not include such spectra, but it is straightforward to extend the model described here to charge > +3 spectra.

[6] Because of conditional independencies in Figure 3(a), $p(\xi)$ factors such that integrating over $\xi$ is done in $O(n)$ time.

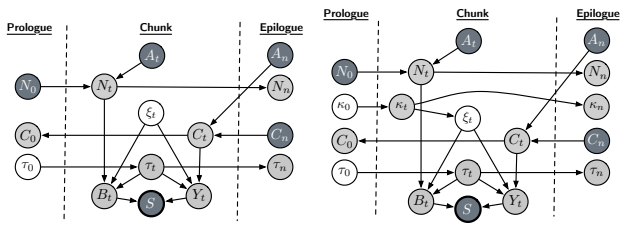

(a) Didea for +3 spectra in database search (Section 3.4)  (b) Didea for multiply-charged spectra (Section 3.5)

Figure 3: Didea models which are used to handle multiply-charged spectra *(a)* $\xi_n$ is the charge of the $b$-ion. *(b)* $\kappa_0$ is a Bernoulli random variable over spectrum charge $\{+2, +3\}$, and is copied in each frame, $\forall t, \kappa_t = \kappa_0$.

product ions exist in the spectrum. In our approach, if many possible peaks are missing from the spectrum, then $p(\xi)$ will get smaller as the disagreement between $\xi$ and the spectrum increases. It is possible for $p(\xi)$ to be infinitesimally small for all but a few values of $\xi$. That is, a high score can be achieved even if only a few $\xi$ have significant probability.

### 3.5 Multiply-charged spectra

Electrospray ionization routinely produces peptide precursor ions where charge $c(s) \in \{+1, +2, +3\}$. When $c(s) = +1$, the spectrum is referred to as *singly-charged*; when $c(s) = +2$, the spectrum is referred to as *doubly-charged*; when $c(s)$ has more than one possible value, we are unable to distinguish which charge with certainty, and the spectrum is referred to as *multiply-charged*.[7]

Existing database search algorithms analyze multiply-charged spectra by doubling the amount of computation (Eng et al., 1994). Each peptide-spectrum match is searched at $c(s) = +2$ and again at $c(s) = +3$, and the higher scoring match is used. If $\Psi_c$ is the search algorithm specialized to charge $c$, replace Equation 2 with

$$p^* = \underset{p \in C(m, P, \delta)}{\operatorname{argmax}} \max_{c \in \{+2, +3\}} \Psi_c(s, p). \qquad (12)$$

We could do the same search over charges with Didea. Evaluating $\theta(\cdot, \cdot)$ using Figure 2 corresponds to $\Psi_{+2}$; using Figure 3(a) corresponds to $\Psi_{+3}$. However, a concern with Equation 12 is that it assumes that the per-charge scoring functions are calibrated,

$$\mathbf{E}_{s \sim S, p \sim P} \left[ \Psi_{+2}(s, p) \right] = \mathbf{E}_{s \sim S, p \sim P} \left[ \Psi_{+3}(s, p) \right]. \quad (13)$$

A multiply-charged spectrum is an expression of uncertainty as to the charge state of the precursor. The mass window in the MS1 step is such that selecting the same peptide, at different charges, in the same MS2 scan, is almost impossible. It is not plausible that a single spectrum represents a mixture of the same peptide at different charges. So while taking the max of a set of charge-specific scoring function $\{\Psi_c\}_c$ allows one to make identification decisions, it provides no measure of uncertainty as to a spectrum's true charge—either charge $+2$ or charge $+3$ is selected, but no distribution over spectrum charge state can be produced.

Thus, rather than taking a maximum over scores for different charge states, we alter Didea to model uncertainty as to the value of $c(s)$. In Figure 3(b) we introduce a Bernoulli random variable $\kappa_0 \in \{+2, +3\}$, which is copied into each chunk frame. Moreover,

$$p(\xi_t \mid \kappa_t) = \begin{cases} p(\xi_t = +1) = 1.0 & \kappa_t = +2, \\ p(\xi_t = +1) = 0.5 & \kappa_t = +3, \end{cases} \qquad (14)$$

We assume no prior knowledge about the distribution of $c(s)$, $p(\kappa_0 = +2) = 0.5$. The rest of the parameters remain unchanged from Section 3.4. When $\kappa_0 = +2$, the model is identical to Figure 2. When $\kappa_0 = +3$, the model is identical to Figure 3(a). Since $\kappa_0$ is uncertain, when $\theta(\cdot, \cdot)$ is computed the score will be an average of both models. Note, however, that the resulting score is not the same as averaging the scores output by the $+2$ and $+3$ model. In each frame $t > 0$, the scoring inference is using an estimate of the spectrum charge based on the product ions considered—i.e., $p(\kappa_t)$ differs at each $t$. We conjecture that $p(\kappa_n \mid a_1, \ldots, a_n)$ can be used as an estimate of the uncertainty in the charge of the peptide-spectrum match chosen by Didea.

## 4 Experiments

Evaluating spectrum identification algorithms is complicated by the lack of test data. One cannot collect realistic spectra from known peptides. Section 4.1 describes how to estimate *absolute ranking curves*, which are the standard approach for comparing spectrum identification algorithms. We benchmark Didea against a panel of spectra from three different organisms (Section 4.2), presenting results in Section 4.3.

### 4.1 Evaluation Without Ground Truth

A correct peptide-spectrum match (PSM) $(p_i, s_i)$ is one where $s_i$ is the result of the fragmentation of peptide $p_i$, and an incorrect PSM is any match that is not correct. Clearly, database search scores vastly more incorrect PSMs than correct ones. Moreover, evaluation is greatly complicated by the lack of ground truth. Con-ceptually, ground truth for a spectrum identification could be generated by feeding a purified peptide $p_*$ into the tandem mass spectrometer. Unfortunately, there is no way of purifying $p_*$ to the point where contaminants are undetectable. Moreover, even if one could run a pure peptide sample through the mass spectrometer, the resulting spectrum $s_*$ would exhibit an unrealistically low level of noise; the test data would not reflect a real shotgun proteomics experiment. The standard solution in shotgun proteomics is to perform an evaluation without ground truth: measure the number of matches at a bound on the false discovery rate (Elias and Gygi, 2007; Käll et al., 2008).

For any peptide-spectrum match $(p_i, s_i)$ with score $v_i$, there are two possibilities: either the peptide $p_i$ generated spectrum $s_i$, or it did not. We refer to these events as hypotheses $H_1$ and $H_0$, respectively. If the scoring function is any good, then we expect the likelihood of $v_i$ to be low under $H_0$, if $(p_i, s_i)$ is the correct match. We can quantify what "low" means here using a hypothesis test,

$$H_0 : v_i \leq c, \qquad H_1 : v_i > c, \qquad (15)$$

where the choice of $c \in \mathbb{R}$ determines the stringency of the test. Informally, the null hypothesis ($H_0$) is that the match is incorrect; the alternate ($H_1$) is that the match is correct.

Characterizing the accuracy at different values of $c$ is greatly simplified by converting the scores to $p$-values, the probability of obtaining a score at least as large as $c$ under the null hypothesis. Denote the cumulative distribution of scores under the null $G_0(c) \triangleq \mathbb{P}(T \leq c \,|\, H_0)$, where $T$ is a random variable representing the PSM score. Given PSM scores $\{v_i\}_i$ the corresponding $p$-values are $p_i = 1 - G_0(v_i)$. Computing $p$-values for a wide range of scores $v_i$ is tantamount to estimating the null distribution. Estimated $p$-values $\{\hat{p}_i\}_i$ are generated by replacing the exact (unknown) null distribution with an estimate $\hat{G}_0(c)$.

For spectrum identification, a widespread approach for estimating the null distribution is *target-decoy search* (Balgley et al., 2007; Elias and Gygi, 2007; Käll et al., 2008), in which a second search is performed against each spectrum using a set of decoy peptides $d \in D$, each of which could not have generated the spectrum. The original peptides $t \in P$ are referred to as targets. Under the constraint that $P \cap D = \emptyset$, decoys are used to generate a sample from the null distribution

$$\bar{v}_i = \max_{d \in D} \Psi(s_i, d), \quad \forall i = 1 \ldots r. \qquad (16)$$

The samples $\{\bar{v}_i\}_i$ are then used in a Monte Carlo estimate of the $p$-values: $\hat{p}_i$ is just the fraction of sampled decoy scores that exceed $v_i$. In our experiments, decoy peptides are generated by randomly permuting the reference proteome, which induces a set of random peptides under a trypsin digest.

Since there are $r$ hypothesis tests (spectra), we can measure the tradeoff between the number of PSMs that are accepted and the stringency of the threshold on the score using false discovery rates. Visually, the tradeoff is represented using an *absolute ranking plot* (Figure 4), where each point on the x-axis is a $q$ value $\tilde{q} \in [0, 1]$, a measure of the false discovery rate (Storey, 2002), and the corresponding value on the y-axis is the number of PSMs accepted at that $q$-value. At $q = 1$, all $r$ identifications are accepted. Because peptide-spectrum matches are often used as input to another estimation task (e.g., protein identification or quantification) our concern is with maximizing performance at small $q$-values, so we only plot $q \in [0, 0.1]$. One method dominates another if its absolute ranking curve is strictly above the absolute ranking curve for the other method.

Absolute ranking measures what the end-user cares about: maximizing the number of spectra identified at their chosen false discovery rate tolerance.

## 4.2 Data sets

The spectra we study are generated from complex proteins samples drawn from three organisms: tryptic digests of a whole-cell lysate of *Saccharomyces cerevisiae* (yeast), whole-cell lysate of *Caenorhabditis elegans* (worm), and liver tissue from *Mus musculus* (mouse). The yeast and worm data are freely available at http://noble.gs.washington.edu/proj/percolator, with a description of the sample preparation procedure, liquid chromatography protocol, and mass spectrometer settings in Käll et al. (2007).

## 4.3 Results

Our primary claim is that Didea outperforms competing systems under absolute ranking, the metric that end-users actually care about. To support our empirical claim, we benchmark Didea against competitors. All algorithms have access to the same peptide database, created by a fully-tryptic *in silico* digest of reference yeast, worm, and mouse proteomes. Candidate peptides are chosen using a mass tolerance of $\delta = 3.0$ Daltons. A static modification is applied to account for cysteine carbidomethylation in all the algorithms. No variable modifications to amino acids are permitted. All spectra are either singly-charged or multiply-charged. All Didea-type scoring functions preprocess the spectra using Equation 8 and use exact inference to compute the PSM score. For Didea, we use the version of the model which integrates over uncertainty in the precursor charge (Section 3.5).

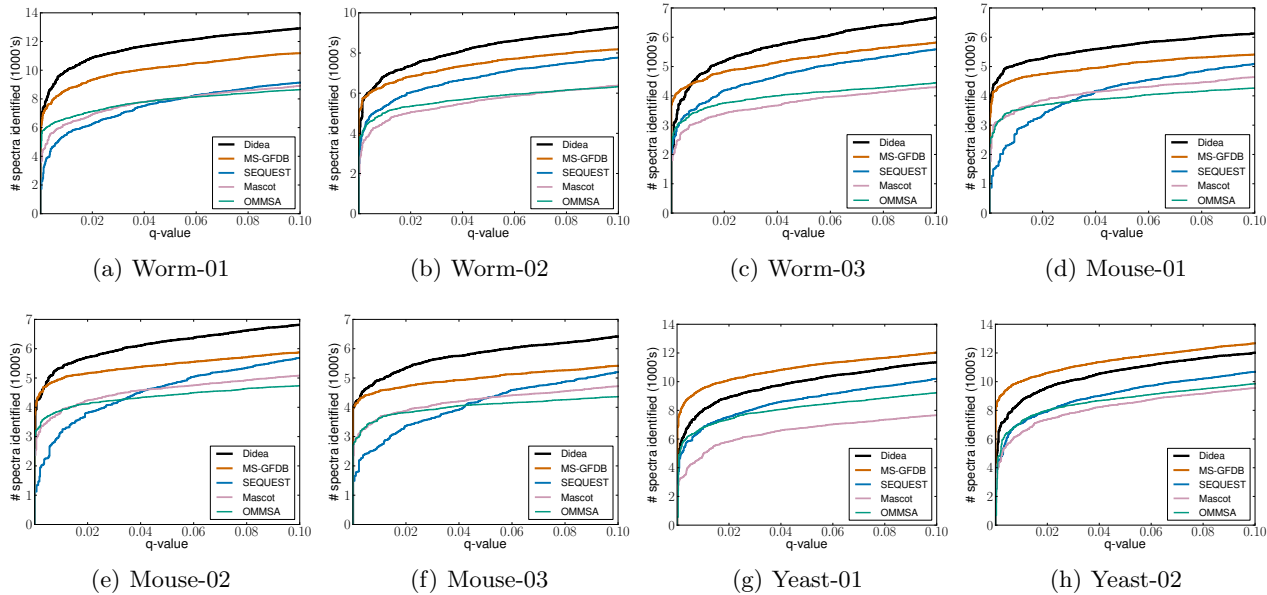

Figure 4: Absolute ranking curves on eight different data sets. Didea significantly outperforms all competitors in (a-f). Didea significantly outperforms all competitors, save MS-GFDB (g-h). If $q = 0.05$, then the expected fraction of false discoveries (incorrect PSMs) in the matches accepted is 0.05. The goal is to maximize the number of spectra identified for $q \in [0, 0.1]$.

The absolute ranking plots for the benchmark are in Figure 4. At an estimated 1% false discovery rate, Didea outperforms MS-GFDB. While we often outperform MS-GFDB, there are cases where it beats Didea (Figures 4(g)-4(h)). Surprisingly, Didea is able to beat MS-GFDB even though it models less of the complexity of peptide fragmentation physics: we use no $a$-ions, nor neutral losses, nor do we account for errors in the $m/z$ measurements in a fragmentation spectrum, like MS-GFDB does. Even more surprising is that Didea achieves high predictive power with only one trained parameter, $\lambda$. Even using a coarse grid search, the chosen value of $\lambda$ works well across a variety of data sets. Training $\lambda$ on a per-organism basis does not significantly increase the absolute ranking curve for Didea. Furthermore, we filter no peaks from the spectrum, even though real spectra are extremely noisy. In contrast, both SEQUEST and MS-GFDB require extensive preprocessing of the spectrum to work well.

All spectrum identification tools make an empirical claim, that their system identifies more spectra at a given $q$-value, or range of $q$-values. We compare favorably on such evaluations, against a wide range of competitors, using a relatively simple model and probabilistic inference.

## 5 Summary and Future Work

We have formulated a scoring function for spectrum identification based on a polynomial-time inference in a dynamic Bayesian network. Didea is significantly more accurate than both SEQUEST and Mascot, the primary tools used for this task in real-world use. Moreover, we outperform even recently developed competitors, like MS-GFDB, on many spectra. We anticipate that adding additional information, e.g., $a$-ions, neutral losses, relationships between ion yield and ion composition, will further improve the accuracy of Didea.

We believe that Didea is suggestive of the promise of applying techniques popular in speech recognition, such as dynamic Bayesian networks, to spectrum identification. For example, natural language processing often uses a compressed representation of a natural language corpora, which can be seen as analogous to compressed representations of a peptide database. Especially exciting would be a variant of Didea which replaces database search with direct Viterbi decoding of the peptide sequence.

**Acknowledgements**: This work was funded by NIH awards R01 GM096306 and P41 GM103533.